\documentclass[12pt]{iopart}
\usepackage{iopams}  
\usepackage{graphics}
\usepackage{epsfig}
\eqnobysec
\begin{document}

\title[The Relativistic Linear Singular Oscillator]{The Relativistic Linear Singular Oscillator}

\author{Shakir M. Nagiyev\dag, Elchin I. Jafarov\dag  
\footnote[3]{To whom correspondence should be addressed
(azhep@physics.ab.az)} and Rizvan M. Imanov\ddag}

\address{\dag\ Institute of Physics, Azerbaijan National Academy of
Sciences, Javid av. 33, 370143, Baku, Azerbaijan}

\address{\ddag\ Physics Department, Ganja State University, 
A. Camil str. 1, 374700, Ganja, Azerbaijan}

\begin{abstract}

Exactly-solvable model of the linear singular oscillator in the relativistic configurational space is considered. We have found wavefunctions and energy spectrum for the model under study. It is shown that they have correct non-relativistic limits.

\end{abstract}


\submitto {\JPA}


\section{Introduction}

In the framework of the non-relativistic quantum mechanics Refs. [1-5] studied in detail the linear singular oscillator, which is described in the configuration representation by the Hamiltonian

\begin{equation}
\label{1}
H_N = - \frac{\hbar ^2}{2m} \frac{d ^2}{dx ^2} + \frac{m \omega ^2} 2 x^2 + \frac g {x^2}.
\end{equation}

The singular oscillator, being one of the rare exactly solvable problems in the non-relativistic quantum mechanics has been extensively used in many applications. There are a lot of quantum mechanical and quantum field theory problems leading to the solution of the Schr\"odinger equation with Hamiltonian (\ref{1}). For example, it served as an initial point in constructing exactly solvable models of interacting $N$-body systems \cite{calogero}. It was also used for the modeling of the diatomic \cite{chumakov} and polyatomic \cite{hartmann} molecules. The interest to the given Hamiltonian recently has increased in connection with its use for the description of spin chains \cite{polychronakos}, quantum Hall effect \cite{frahm}, fractional statistics and anyons \cite{leinaas}. 

In the case of a constant frequency $\omega$, a complete set of orthonormalized eigenfunctions of Hamiltonian (\ref{1}) in the interval $0<x< \infty$ can be choosen in the form (see, for example \cite{dodonov1})

\begin{equation}
\label{2}
\fl \qquad \psi _n^{nonrel}(x)=\left[ 2\left( \frac{m\omega }\hbar \right) ^{d+1}\frac{n!}{\Gamma (d+n+1)}\right] ^{\frac 12}x^{d+\frac 12}e^{-\frac{m\omega }{2\hbar }x^2}L_n^d\left( \frac{m\omega }\hbar x^2\right),
\end{equation}
where $d=\frac 12 \sqrt{1+\frac{8mg}{\hbar ^2}}$, and $L_n^d(y)$ are the associated Laguerre polynomials. The corresponding eigenvalues of $H_N$ (\ref{1}) have the form

\begin{equation}
\label{3}
E_n ^{nonrel} = \hbar \omega (2n+d+1), \quad n=0,1,2,3, \dots \; .
\end{equation}

The purpose of this paper is to construct and investigate a relativistic exactly solvable model of the quantum linear singular oscillator (\ref{1}), which can be applied for study relativistic physical systems as well as systems on a lattice.

Our construction is based on the version of the relativistic quantum mechanics, which was developed in several papers and applied for solution of a lot of problems in particle physics [11-24]. Although this version of the relativistic quantum mechanics is closely analogous to the non-relativistic quantum mechanics, its essential characteristic is that the relative motion wavefunction satisfies a finite-difference equation with the step equal to the Compton wavelength of the particle, $\lambda=\hbar/mc$. For example, in the case of a local quasipotential of interaction $V\left( \vec r \right) $ the equation for the wavefunction of two scalar particles with equal masses has the form

\begin{equation}
\label{4}
\left[ H_0+V\left( \vec r \right) \right] \psi (\vec r)=E\psi (\vec r) \; ,
\end{equation}
where the finite-difference operator $H_0$ is a relativistic free Hamiltonian

\begin{equation}
\label{5}
\fl \qquad H_0=mc^2\left[ \cosh \left( i\lambda \partial _r\right) + \frac{i \lambda}r\sinh \left( i \lambda \partial _r\right) + \frac{  {\vec L}^2}{2\left( mcr \right) ^2}\exp \left( i\lambda \partial _r\right) \right]  \; ,
\end{equation}
and $\vec L^2$ is the square of the angular momentum operator and $\partial _r \equiv \frac \partial {\partial r}$. The technique of difference differentiation was developed and analogues of the important functions of the continuous analysis were obtained to fit the relativistic quantum mechanics, based on Eq. \eref{4} \cite{kadyshevsky2a, kadyshevsky2}.

Unlike non-relativistic quantum mechanics, because of presence of the finite-difference operator $e^{i\lambda \partial_r}$ in presented relativistic quantum mechanics both the Hamiltonian and the quasipotential contain the imaginary terms. However, they are hermitian operators with the respect to scalar product $\int \psi^* \varphi d \vec r$, where $\psi \left(\vec r \right)$ and $\varphi \left( \vec r \right)$ are the square-integrable functions.

We note that, there is a regular method for construction of the quasipotential in the framework of the field-theoretical formalism. However, it can be also introduced phenomenologically. 

The space of vectors $\vec r$ is called the relativistic configurational space or $\vec r$-space. The concept of the $\vec r$-space has been introduced for the first time in the context of the quasipotential approach to the relativistic two-body problem \cite{kadyshevsky1}.

The quasipotential equations for the relativistic scattering amplitude and the wave function $\psi \left( \vec p \right)$ in the momentum space have the form \cite{kadyshevsky3, kadyshevsky4}

\begin{eqnarray}
\label{5a}
\fl \qquad A\left( \vec p,\vec q\right) =\frac m{4\pi }V\left( \vec p,\vec q;E_q\right) +\frac 1{\left( 2\pi \right) ^3}\int V\left( \vec p,\vec k;E_q\right) G_q\left(  k\right) A\left( \vec k,\vec q\right) d\Omega _k \; , \\
\label{6a}
\fl \qquad \psi \left( \vec p\right) =\left( 2\pi \right) ^3\delta \left( \vec p\left( -\right) \vec q\right) +\frac 1{\left( 2\pi \right) ^3}G_q\left( p\right) \int V\left( \vec p,\vec k;E_q\right) \psi \left( \vec k\right) d\Omega _k \; ,
\end{eqnarray}
where

\begin{eqnarray}
\label{7a}
G_q\left( p\right) =\frac 1{E_q-E_p+i0},\quad \delta \left( \vec p\left( -\right) \vec q\right) \equiv \sqrt{1+\frac{\vec q^2}{m^2c^2}}\delta \left( \vec p-\vec q\right) , \\
d\Omega _k=\frac{d\vec k}{\sqrt{1+\frac{\vec k^2}{m^2c^2}}},\quad E_q=\sqrt{\vec q^2c^2+m^2c^4}, \nonumber
\end{eqnarray}
and $V\left(\vec p, \vec k; E_q \right)$ is the quasipotential.

The integration in \eref{5a} and \eref{6a} is carried over the mass shell of the particle with mass $m$, i.e. over the upper sheet of the hyperboloid ${p_0}^2-{\vec p}^2=m^2c^2$, which from the geometrical point of view realizes the three-dimensional Lobachevsky space. The group of motions of this space is the Lorentz group $SO(3,1)$.

Equations \eref{5a} and \eref{6a} have the absolute character with respect to the geometry of the momentum space, i.e., formally they don't differ from the non-relativistic Lippmann-Schwinger and Schr\"odinger equations. We can derive Eqs. \eref{5a} and \eref{6a} substituting the relativistic (non-Euclidean) expressions for the energy, volume element, and $\delta$-function by their non-relativistic (Euclidean) analogues:

\begin{eqnarray}
\label{8a}
E_q=\frac{q^2}{2m} \rightarrow E_q = \sqrt{{\vec q}^2 c^2 +m^2 c^4} \; , \nonumber \\
d \vec k \rightarrow d \Omega _k = \frac{d\vec k}{\sqrt{1+\frac{{\vec k}^2}{m^2c^2}}}\;, \\
\delta\left(\vec p - \vec q \right) \rightarrow \delta\left(\vec p (-) \vec q \right). \nonumber
\end{eqnarray}

As a consequence of this geometrical treatment, the application of the Fourier transformation on the Lorentz group becomes natural instead of usual one. In this case the relativistic configurational $\vec r$-space conseption arises.

Transition to relativistic configurational $\vec r$-representation

\begin{equation}
\label{9a}
\psi \left( \vec r\right) =\frac 1{\left( 2\pi \hbar \right) ^{\frac 32}}\int \xi \left( \vec p,\vec r\right) \psi \left( \vec p\right) d\Omega _p
\end{equation}
is performed by the use of expansion on the matrix elements of the principal series of the unitary irreducible representations of the Lorentz group:

\begin{eqnarray}
\label{10a}
\xi \left( \vec p,\vec r\right) =\left( \frac{p_0-\vec p\vec n}{mc}\right) ^{-1-ir/\lambda } \; ,  \\
\vec r=r\vec n,\quad 0\leq r<\infty ,\nonumber \\ \vec n=\left( \sin \theta \cos \varphi ,\sin \theta \sin \varphi ,\cos \theta \right) ,\; p_0=\sqrt{\vec p^2+m^2c^2}. \nonumber
\end{eqnarray}

The quantity $r$ is relativistic invariant and is connected with the eigenvalues of the Casimir operator $\hat C = \vec N ^2- \vec L ^2$ in the following way

\begin{equation}
\label{11a}
C=\lambda  ^2 + r ^2 \; ,
\end{equation}
where $\vec L$ and $\vec N$ are the rotation and boost generators.

It is easy to verify that the function (the relativistic "plane wave") \eref{10a} obeys the finite-difference Schr\"odinger equation

\begin{equation}
\label{12a}
\left(H_0-E_p \right) \xi \left(\vec p, \vec r \right) = 0 \; .
\end{equation}

The relativistic plane waves form a complete and orthogonal system of functions in the momentum Lobachevsky space \cite{shapiro}.

If we perform the relativistic Fourier transformation \eref{9a} in Eq. \eref{6a}, we arrive to the finite-difference Schr\"odinger equation \eref{4} with the local (in general case, non-local) potential in the relativistic $\vec r$ space.

In the relativistic $\vec r$ space the Euclidean geometry is realized and, in particular, there exist a momentum operator in the relativistic configurational $\vec r$ representation \cite{kadyshevsky2}

\begin{equation}
\label{13a}
\hat {\vec p} = - \vec n \left( e ^{i \lambda \partial _r} - H_0 \right) - \vec m \frac{1}{r} e ^{i \lambda \partial _r} \; ,
\end{equation}
where a three-dimensional vector $\vec m$ has the following components \cite{nagiyev1}:

\begin{eqnarray}
m_1=i\left( \cos \varphi \cos \theta \frac \partial {\partial \theta }-\frac{\sin \varphi }{\sin \theta }\frac \partial {\partial \varphi }\right) \; , \nonumber \\
m_2=i\left( \sin \varphi \cos \theta \frac \partial {\partial \theta }-\frac{\cos \varphi }{\sin \theta }\frac \partial {\partial \varphi }\right) \; , \nonumber \\
m_3=-i\sin \theta \frac \partial {\partial \theta } \; .
\end{eqnarray}

The components of \eref{13a} and free Hamiltonian obey the following commutation relations:

\begin{equation*}
\left[ \hat p_i,\hat p_j\right] =\left[ \hat p_i,H_0\right] =0,\quad i,j=1,2,3 \; .
\end{equation*}

The relativistic plane wave is the eigenfunction of the operator $\hat {\vec p}$:

\begin{equation}
\label{14a}
\hat {\vec p} \xi \left(\vec p, \vec r \right) = \vec p \xi \left(\vec p, \vec r \right) \; .
\end{equation}

This means that \eref{10a} describes the free relativistic motion with definite energy and momentum.

In the non-relativistic limit we come to the usual three-dimensional configurational space and relativistic plane wave (\ref{10a}) goes over into the Euclidean plane wave, i.e.

\begin{equation}
\label{8}
\lim _{c \rightarrow \infty} \xi \left( \vec p, \vec r \right) = e ^{i \vec p \vec r / \hbar} \; .
\end{equation}

Note that all the important exactly solvable cases of non-relativistic quantum mechanics (potential well, Coulomb potential, harmonic oscillator etc.) are exactly solvable for the case of Eq. (\ref{4}), too.

\section{Relativistic Quantum Mechanics: The One-Dimensional Case}

In the one-dimensional case the relativistic plane wave takes a form \cite{atakishiyev1}

\begin{equation}
\label{9}
\xi \left( p,x\right) =\left( \frac{p_0-p}{mc}\right) ^{-ix/\lambda }  =\left( \frac{p_0+p}{mc}\right) ^{ix/\lambda } \; ,
\end{equation}
or, in hyperpolar coordinates

\begin{equation}
\label{10}
p_0=mc \cosh \chi \; , \quad p=mc \sinh \chi \; ,
\end{equation}
we have

\begin{equation}
\label{11}
\xi (p,x) = e^{ix \chi / \lambda } \; , 
\end{equation}
where $\chi= \ln \left( \frac{p_0+p}{mc} \right)$ is rapidity.

The one-dimensional plane waves obey the completeness and orthogonality conditions

\numparts
\begin{eqnarray}
\label{12}
\frac {1}{2\pi \hbar }\int\limits_{-\infty }^\infty \xi \left( p,x\right) \xi ^{*}\left( p,x^{\prime }\right) d\Omega _p=\delta \left( x-x^{\prime }\right) \; , \\
\frac {1}{2\pi \hbar }\int\limits_{-\infty }^\infty \xi \left( p,x\right) \xi ^{*}\left( p^{\prime },x\right) dx= \delta \left( p (-) p^{\prime }\right) = \delta \left(mc \left( \chi -\chi ^{\prime }\right)\right) \; ,
\end{eqnarray}
\endnumparts
where $d \Omega _p = mc\frac {d p}{p_0} = mc d \chi$ is the invariant volume element in the one-dimensional Lobachevsky space, realized on the upper sheet of the hyperbola $p_0^2-p^2=m^2c^2, \; p_0>0$.

The free Hamiltonian and momentum operators are finite-difference operators

\begin{equation}
\label{13}
\hat H _0 =mc ^2 \cosh \left( i \lambda \partial _x \right) \; , \quad \hat p = -mc \sinh \left( i \lambda \partial _x \right) \; .
\end{equation}

Plane wave (\ref{9}) obeys the free relativistic finite-difference Schr\"odinger equation

\begin{equation}
\label{14}
\left( \hat H_0-E_p\right) \xi \left( p,x\right) =0 \; , \quad E_p=cp_0=c\sqrt{p^2+m^2c^2} \; ,
\end{equation}
and
\begin{equation*}
\hat p \xi \left( p, x \right) = p \xi \left( p, x \right) \; .
\end{equation*}

\section{The Finite-Difference Relativistic Model of the Linear Singular Oscillator}

We consider a model of the relativistic linear singular oscillator, which corresponds to the following interaction potential

\begin{equation}
\label{15}
V\left( x\right) =\left[ \frac 12m\omega ^2x\left(x+i\lambda \right)+\frac g{x\left(x+i\lambda \right)}\right] e^{i\lambda \partial _x} \; ,
\end{equation}
where $g$ is a real quantity (we will assign a restriction for the values of the parameter $g$ below).

For $g=0$ it coincides with quasipotential of the relativistic linear oscillator studied in detail in \cite{atakishiyev1}.

Let us note that in contrast to the case of the Coulomb potential \cite{freeman, nagiyev1, nagiyev2}, which can be calculated as an input of the one-photon exchange, the relativistic generalization of the oscillator or singular oscillator potential is not uniquely defined. Therefore, for construction of the quasipotential \eref{15} we proceed from the following requirements for the quasipotential: a) exact solubility; b) the correct non-relativistic limit; c) exsistence of the dynamical symmetry.

The operator \eref{15} is hermitian with respect to a scalar product

\begin{equation}
\label{15a}
\left( \psi ,\varphi \right) =\int\limits_{-\infty }^\infty \psi ^{*}(x)\varphi (x)dx \; ,
\end{equation}
i.e. $\left( V\psi ,\varphi \right) =\left( \psi ,V\varphi \right)$. Here functions $\psi (x)$ and $\varphi(x)$ vanish at $x= \pm \infty$ together with all their derivatives. In this connection, we note that, the hermitian conjugate of operator $A=f(x)e^{\gamma \partial _x}$ with the respect to the scalar product \eref{15a} has a following form:

\begin{equation*}
A^+=e^{-\gamma^* \partial _x}f^*(x) \; ,
\end{equation*}
where $f(x)$ is some complex function.

It is to be emphasized that potential (\ref{15}) possesses the correct non-relativistic limit, i.e.

\begin{equation*}
\lim _{c \rightarrow \infty} V(x) = \frac 12 m \omega ^2 x^2 + \frac g {x^2} \; .
\end{equation*}

A relativistic singular oscillator is described by the following finite-difference equation

\begin{equation}
\label{16}
\fl \; \left[ mc^2\cosh i \lambda \partial _x+\frac 12m\omega ^2x(x+i\lambda )e^{i\lambda \partial _x}+\frac g{x\left(x+i\lambda \right)}e^{i\lambda \partial _x}\right] \psi (x)=E\psi (x),
\end{equation}
with the boundary conditions for the wavefunction $\psi (0)=0$ and $\psi (\infty)=0$.

We shall confine ourselves to the interval $0 \leq x < \infty$.

In in terms of dimensionless variable $\rho =  \frac{x}{\lambda}$ and parameters $\omega _0 = \frac {\hbar \omega}{mc^2}$, $g _0 =\frac{mg}{\hbar ^2}$ the equation (\ref{16}) takes a form:

\begin{equation}
\label{17}
\left[ \cosh i\partial _\rho +\frac 12\omega _0^2\rho ^{(2)}e^{i\partial _\rho }+\frac{g_0}{\rho ^{(2)}}e^{i\partial _\rho }\right] \psi (\rho )=\frac E{mc^2}\psi (\rho ) \; ,
\end{equation}
where $\rho ^{(2)}= \rho (\rho + i)$.

To solve the equation \eref{17} we choose $\psi(\rho)$ as

\begin{equation}
\label{18}
\psi (\rho )=c(-\rho )^{(\alpha )}\omega _0^{i\rho }\Gamma \left( \nu +i\rho \right) \Omega (\rho )\equiv c(-\rho )^{(\alpha )}M(\rho )\Omega (\rho ) \; ,
\end{equation}
where $\alpha$ and $\nu $ are arbitrary constant parameters, which will be defined below.

Functions

\begin{equation}
\label{19}
(-\rho )^{(\alpha )}=i^\alpha \frac{\Gamma \left( i\rho +\alpha \right) }{\Gamma \left( i\rho \right) }\quad {\rm and} \quad M(\rho )=\omega _0^{i\rho }\Gamma \left( \nu +i\rho \right)
\end{equation}
are connected with the behaviour of the wavefunction $\psi (\rho)$ at points $\rho =0 $ and $\rho = \infty$, respectively.

Inserting (\ref{18}) into (\ref{17}), we obtain:

\begin{equation}
\label{20}
\fl \qquad \left[ \left( \alpha +i\rho \right) \left( \nu +i\rho \right) e^{-i\partial _\rho }-\frac{\omega _0^2\left( \rho ^{(2)}\right) ^2+\rho ^{(2)}+2g_0}{\omega _0^2\left( \alpha -1+i\rho \right) \left( \nu -1+i\rho \right) }e^{i\partial _\rho }\right] \Omega (\rho )=2i \epsilon \rho \Omega (\rho ) \; ,
\end{equation}
where $\epsilon =E/ \hbar \omega$ is dimensionless energy.

Now we choose constant parameters $\alpha$ and $\nu$ in such a way, that they satisfy the following relation:

\begin{equation}
\label{21}
\frac{\omega _0^2\left( \rho ^{(2)}\right) ^2+\rho ^{(2)}+2g_0}{\omega _0^2\left( \alpha -1+i\rho \right) \left( \nu -1+i\rho \right) }=\left( \alpha -i\rho \right) \left( \nu -i\rho \right) \; .
\end{equation}

Hence we get

\numparts
\begin{eqnarray}
\label{22}
\alpha =\frac 12+\frac 12\sqrt{1+\frac 2{\omega _0^2}\left( 1-\sqrt{1-8g_0\omega _0^2}\right) } \; , \\
\nu =\frac 12+\frac 12\sqrt{1+\frac 2{\omega _0^2}\left( 1+\sqrt{1-8g_0\omega _0^2}\right) } \; .
\end{eqnarray}
\endnumparts

Then the function $\Omega (\rho )$ will satisfy the difference equation

\begin{equation}
\label{23}
\left[ \left( \alpha +i\rho \right) \left( \nu +i\rho \right) e^{-i\partial _\rho }-\left( \alpha -i\rho \right) \left( \nu -i\rho \right) e^{i\partial _\rho }\right] \Omega (\rho )=2i \epsilon \rho \Omega (\rho ) \; .
\end{equation}

By substitution of the $\Omega (\rho)$ function expansion as following power series

\begin{equation}
\label{24}
\Omega (\rho )=\sum\limits_{k=0}^\infty e_k(i\rho )^k
\end{equation}
into \eref{23} one finds that the coefficients at odd degrees of $i \rho$ become zero (i.e., all $e_{2k+1}=0, \; k=0,1,2,3, \dots$), but the coefficients at even degrees of $i \rho$ satisfy following recurrence relation:

\begin{equation}
\label{25}
\left( \epsilon -\alpha -\nu -2j\right) e_{2j}=\sum\limits_{k=j+1}^\infty \left[ \alpha \nu C_{2k}^{2j+1}+(\alpha +\nu )C_{2k}^{2j}+C_{2k}^{2j-1}\right] e_{2k} \; ,
\end{equation}
where $C_n^m$ are binomial coefficients. From \eref{25} it follows that the power series \eref{24} will be terminated at the term $e_{2n} (i \rho)^ {2n}$ if condition $\epsilon \equiv \epsilon _n = 2n + \alpha +\nu, \; n=0,1,2,3, \dots$ holds. This gives following quantization rule for the energy spectrum for relativistic singular oscillator \eref{15}:

\begin{equation}
\label{26}
E _n = \hbar \omega \epsilon _n = \hbar \omega (2n+ \alpha + \nu), \; n=0,1,2,3, \dots \; .
\end{equation}

Hence in the case of \eref{26} solutions of the equation \eref{23} coincide with the continuous dual Hahn polynomials

\begin{equation}
\label{27}
\Omega (\rho) \equiv \Omega _n (\rho) = S_n \left(\rho ^2; \alpha, \nu, \frac12 \right) \; ,
\end{equation}
defined with relation \cite{koekoek} 

\begin{equation}
\label{28}
\fl \qquad S_n\left( x^2;a,b,c\right) =\left( a+b\right) _n\left( a+c\right) _n\ _3F_2\left( \begin{array}{c|c} {\begin{array}{c} -n,\quad a+ix,\quad a-ix \\ a+b,\quad a+c \end{array}}  & 1 \end{array} \right) \; ,
\end{equation}
where $(a)_n=a(a+1) \cdots (a+n-1) = \Gamma (a+n) / \Gamma (a)$ is Pochhammer symbol.

Continuous dual Hahn polynomials \eref{28} satisfy the three-term recurrence relation \cite{koekoek}

\begin{equation}
\label{29}
\left(A_n +C_n -a^2 - x^2 \right) \tilde S_n\left( x^2\right)  = A_n \tilde S_{n+1} \left( x^2\right) +C_n \tilde S_{n-1} \left( x^2\right) \; ,
\end{equation}
where $A_n =(n+a+b)(n+a+c), \; C_n = n(n+b+c-1)$ and

\begin{equation*}
\tilde S_n \left( x^2\right) = \frac{S_n\left( x^2;a,b,c\right)}{(a+b)_n (a+c)_n} \; .
\end{equation*}

Hence normalized wavefunctions for the stationary states of the relativistic singular linear oscillator have the following form:

\begin{eqnarray}
\label{30}
\psi _n(\rho )=c_n(-\rho )^{\left( \alpha \right)} \omega _0^{i\rho }\Gamma \left( \nu +i\rho \right) S_n\left( \rho ^2;\alpha ,\nu ,\frac 12\right) \; , \\
c_n=\sqrt{\frac 2{\Gamma \left( n+\alpha +\nu \right) \Gamma \left( n+\alpha +\frac 12\right) \Gamma \left( n+\nu +\frac 12\right) n!}} \; . \nonumber
\end{eqnarray}
\begin{figure}[h!]
\begin{minipage}[b]{0.5\linewidth} 
\epsfig{file=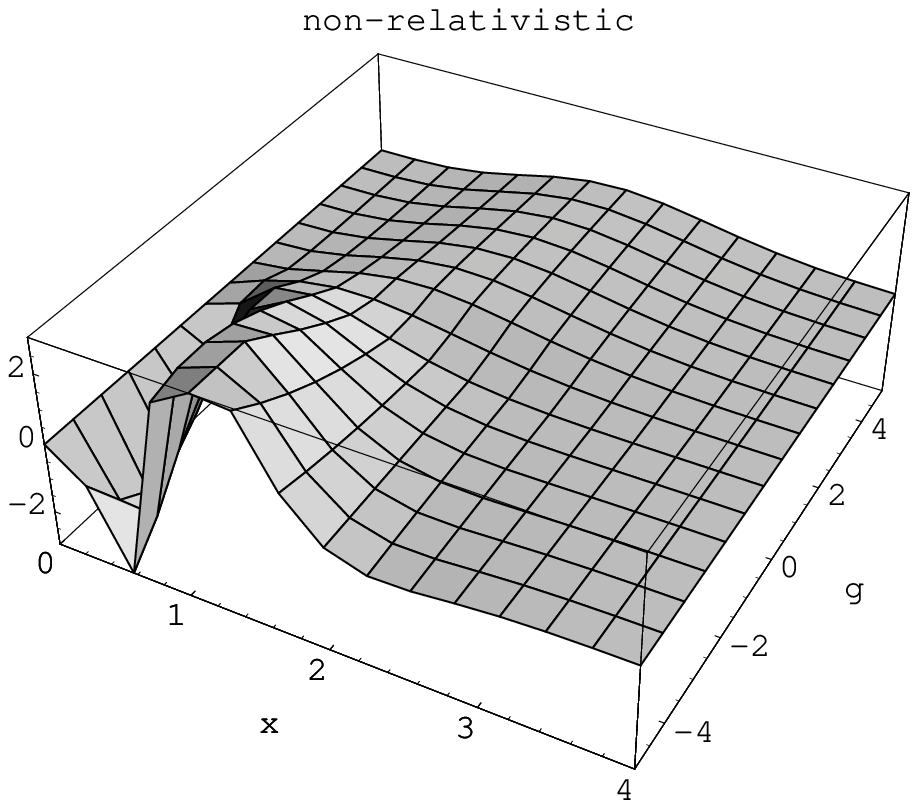,height=5.59cm,width=6.64cm}
\end{minipage}
\hspace{0.5cm} 
\begin{minipage}[b]{0.5\linewidth}
\epsfig{file=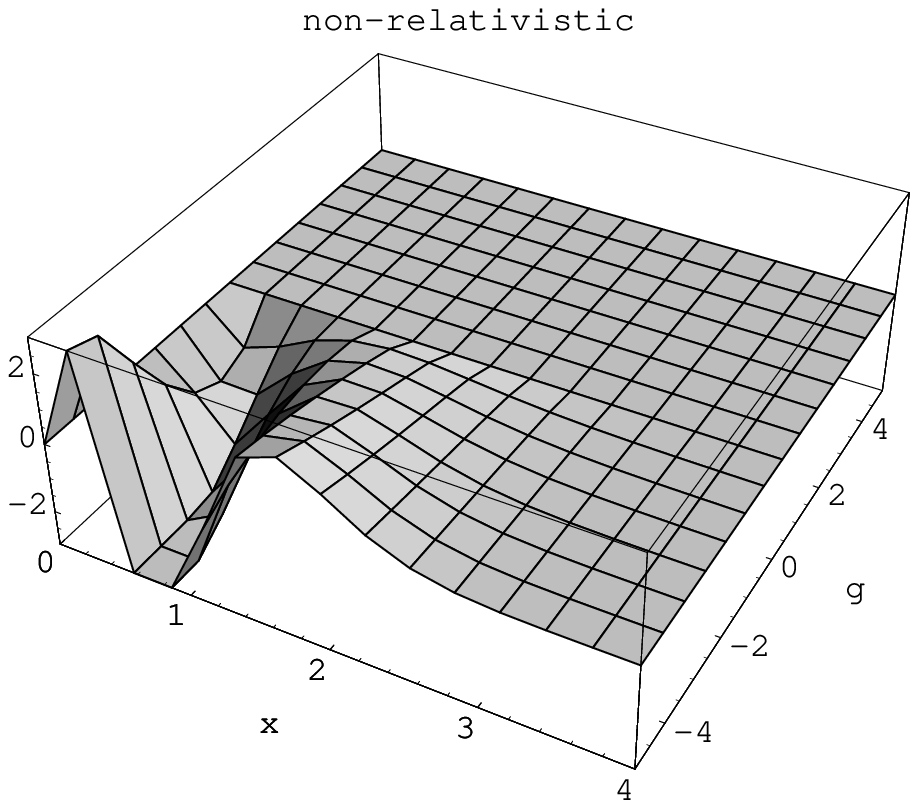,height=5.59cm,width=6.64cm}
\end{minipage}
\begin{minipage}[b]{0.5\linewidth} 
\epsfig{file=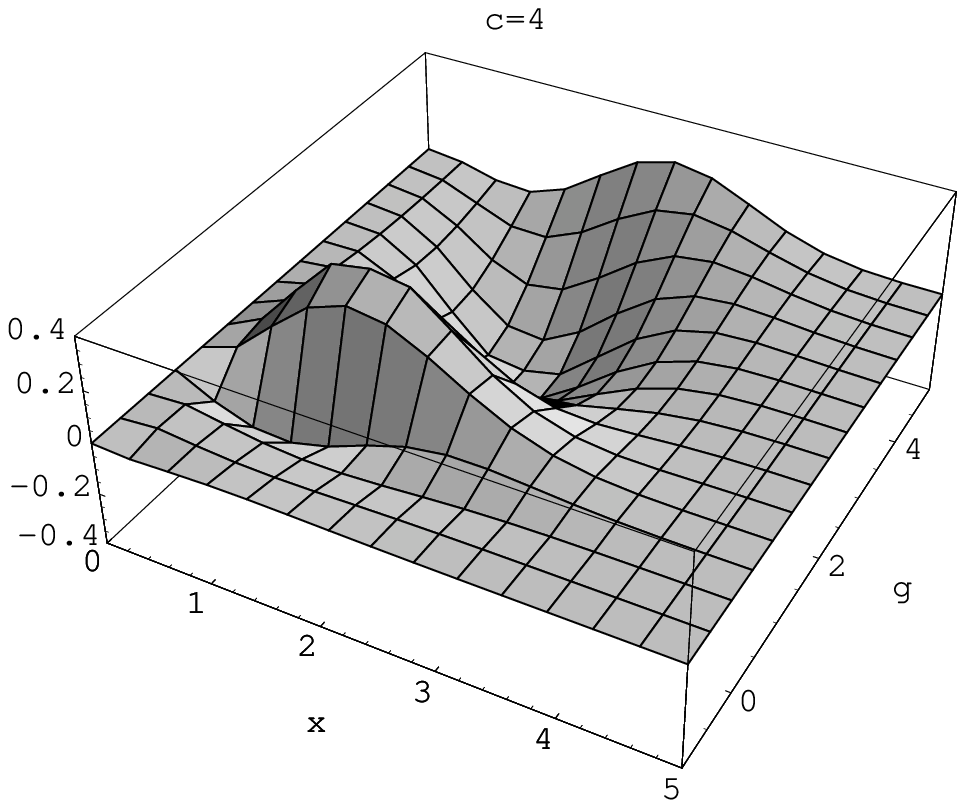,height=5.59cm,width=6.64cm}
\end{minipage}
\hspace{0.5cm} 
\begin{minipage}[b]{0.5\linewidth}
\epsfig{file=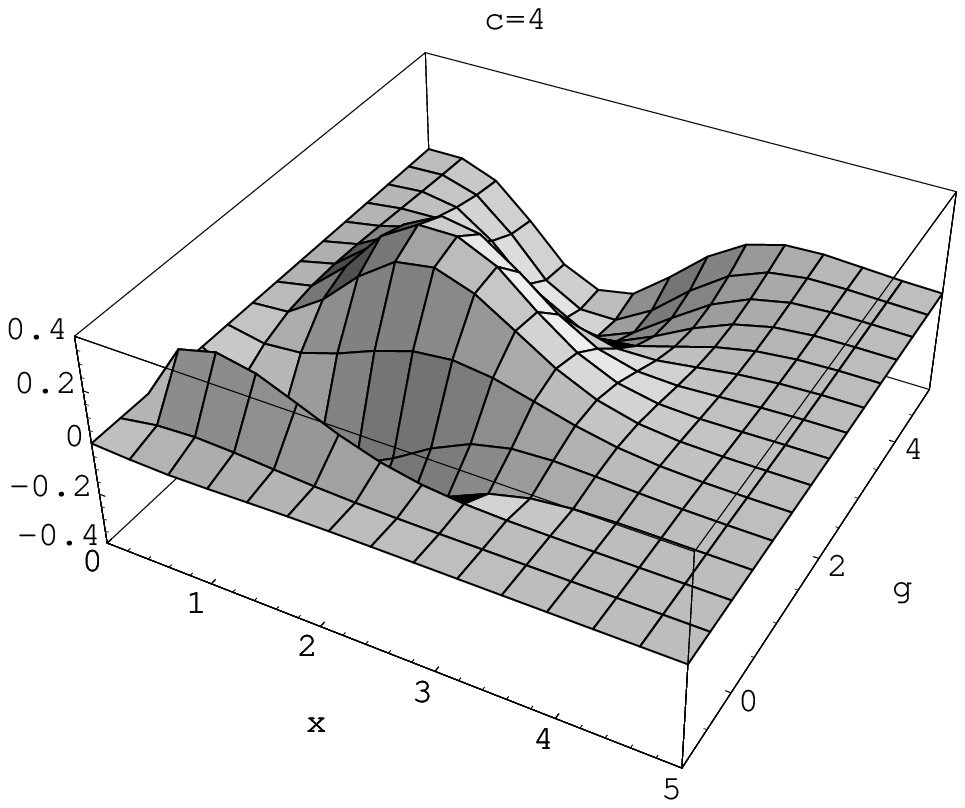,height=5.59cm,width=6.64cm}
\end{minipage}
\begin{minipage}[b]{0.5\linewidth} 
\epsfig{file=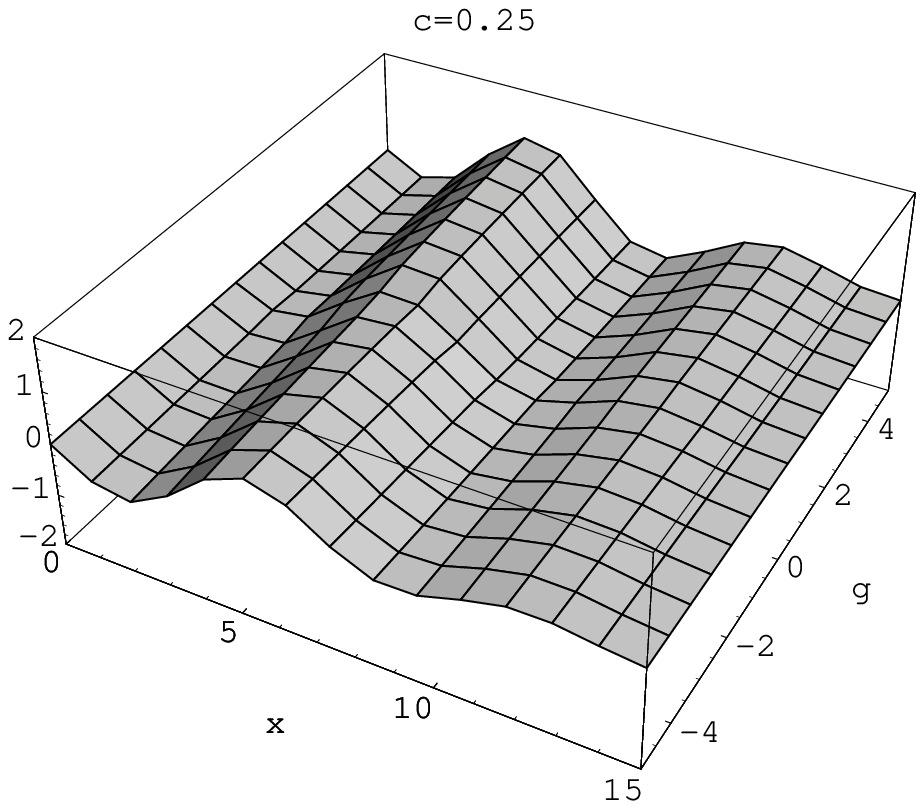,height=5.59cm,width=6.64cm}
\end{minipage}
\hspace{0.5cm} 
\begin{minipage}[b]{0.5\linewidth}
\epsfig{file=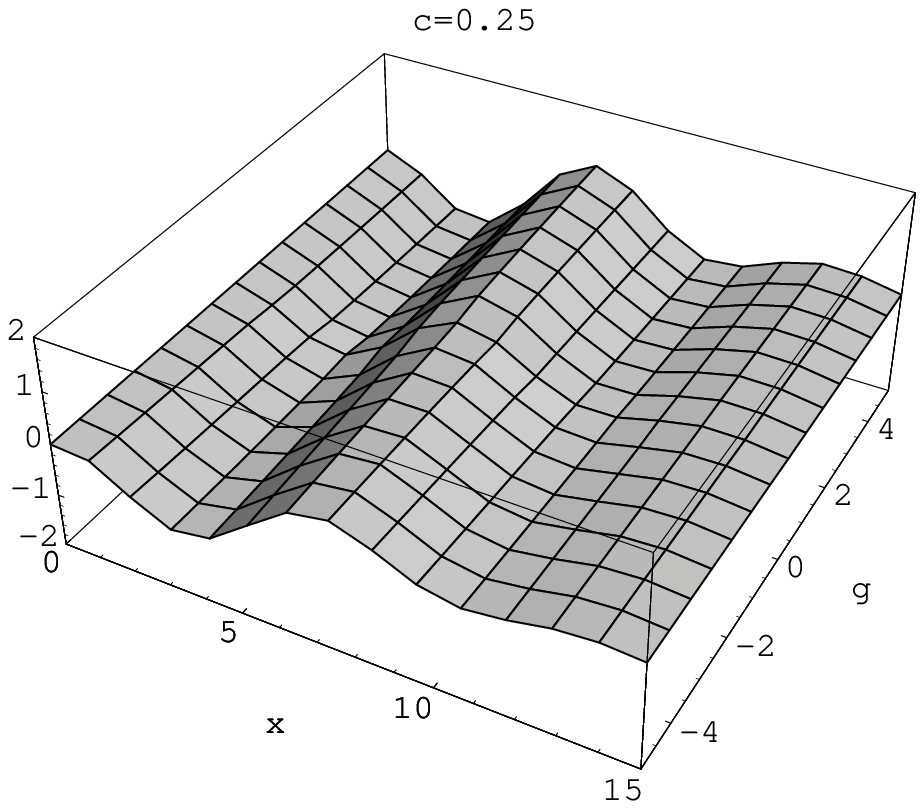,height=5.59cm,width=6.64cm}
\end{minipage}
\caption{\label{fig1} The behaviour of the ground state wave function of the relativistic linear singular oscillator \eref{30} for values of speed of light $c= \infty$ (the non-relativistic case), $4$ and $0.25$ ($m=\omega=\hbar=1$). Real parts are shown in the left plots, imaginary parts in the right plots.}
\end{figure}

Wavefunctions \eref{30} are orthonormalized as follows:

\begin{equation*}
\int\limits_0^\infty \psi _n(\rho )\psi _m^{*}(\rho )d\rho =\delta _{nm} \; .
\end{equation*}

One can verify that the difference Hamiltonian of equation \eref{16}

\begin{equation}
\label{31a}
H=mc^2 \left[ a^+ a^- + \omega _0 \left( \alpha + \nu \right) \right]
\end{equation}
may be factorized in terms of the operators, having the form

\numparts
\begin{eqnarray}
\label{32b}
a^-= \frac {1}{\sqrt{2}} \left[e ^ {-\frac {i}{2} \partial _\rho} - \omega _0 e ^ {\frac {i}{2} \partial _\rho} \left( \nu +i \rho\right) (1+ \frac {\alpha}{i \rho}) \right] \;, \\
\label{32c}
a^+= \frac {1}{\sqrt{2}} \left[e ^ {-\frac i2 \partial _\rho} - \omega _0  \left( \nu -i \rho \right) (1- \frac {\alpha}{i \rho}) e ^ {\frac i2 \partial _\rho} \right] \;.
\end{eqnarray}
\endnumparts

They are the pair of hermitian conjugate operators. Using $a^-$ and $a^+$, one can construct a dynamical symmetry algebra of the relativistic linear oscillator \eref{16}.

\begin{figure}[h!]
\begin{minipage}[b]{0.5\linewidth} 
\epsfig{file=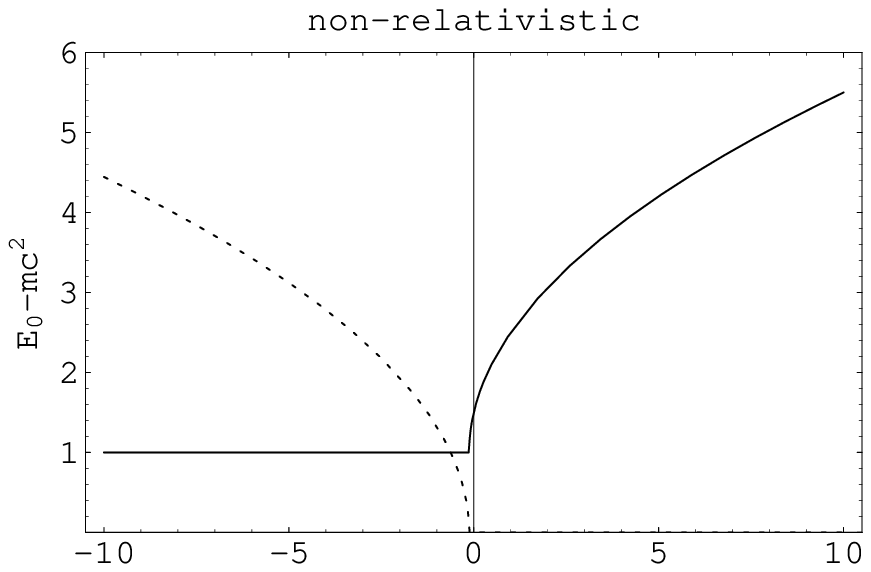,height=5.52cm,width=7.93cm}
\end{minipage}
\hspace{0.5cm} 
\begin{minipage}[b]{0.5\linewidth}
\epsfig{file=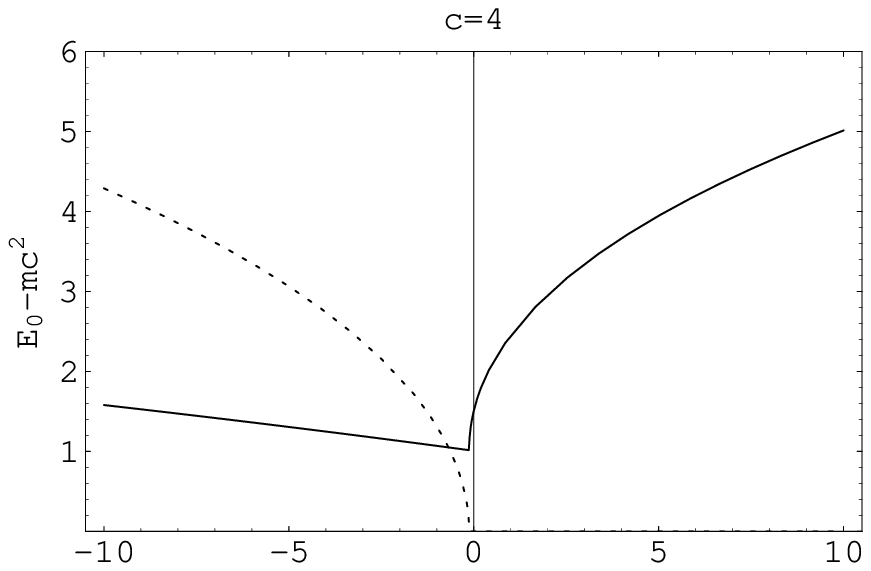,height=5.52cm,width=7.93cm}
\end{minipage}
\begin{minipage}[b]{0.5\linewidth} 
\epsfig{file=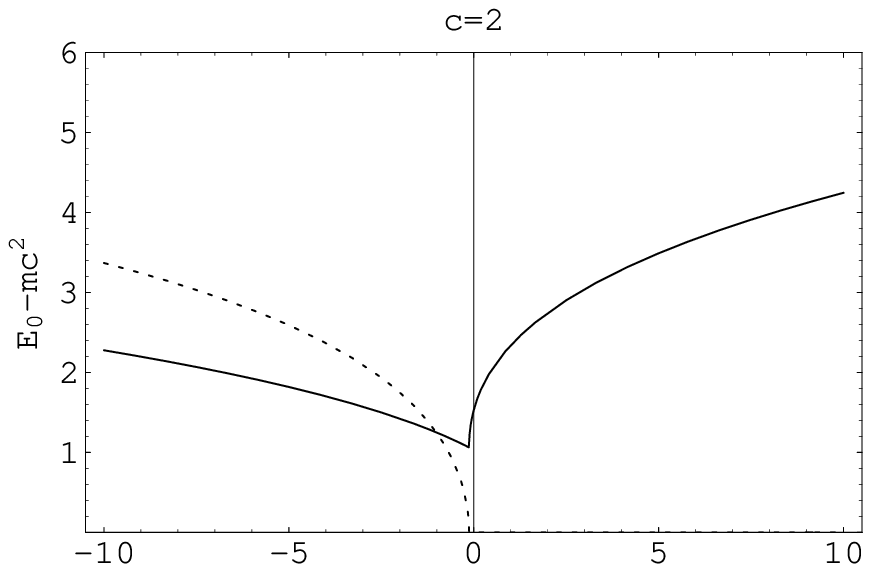,height=5.52cm,width=7.93cm}
\end{minipage}
\hspace{0.5cm} 
\begin{minipage}[b]{0.5\linewidth}
\epsfig{file=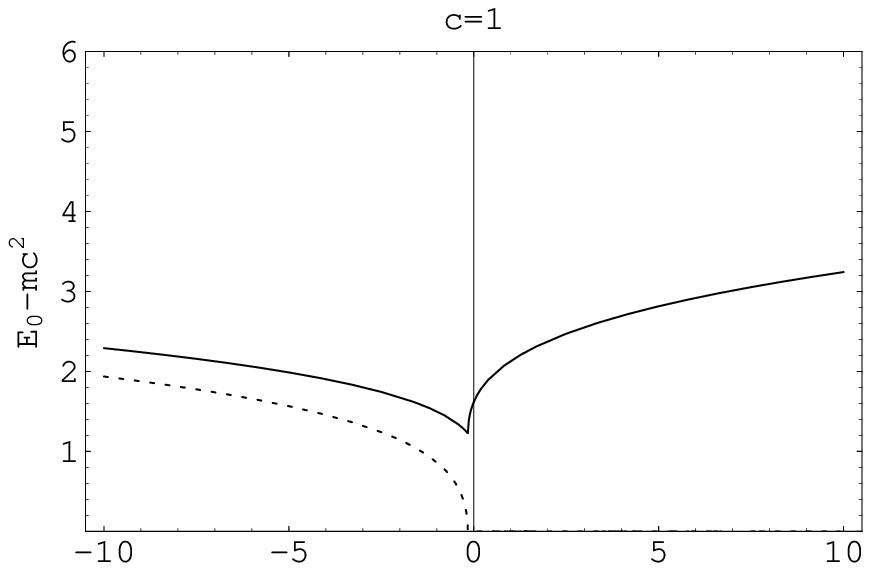,height=5.52cm,width=7.93cm}
\end{minipage}
\begin{minipage}[b]{0.5\linewidth} 
\epsfig{file=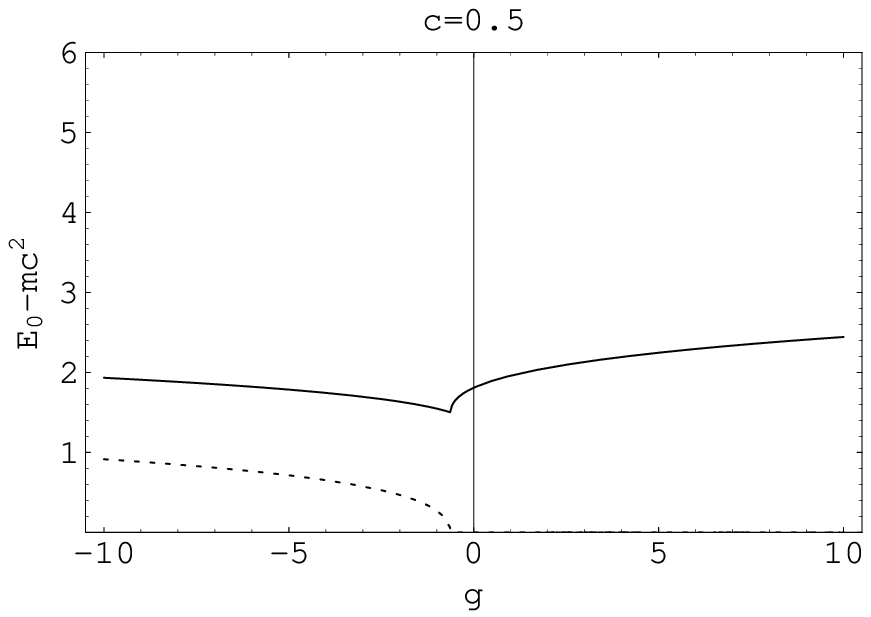,height=5.52cm,width=7.93cm}
\end{minipage}
\hspace{0.5cm} 
\begin{minipage}[b]{0.5\linewidth}
\epsfig{file=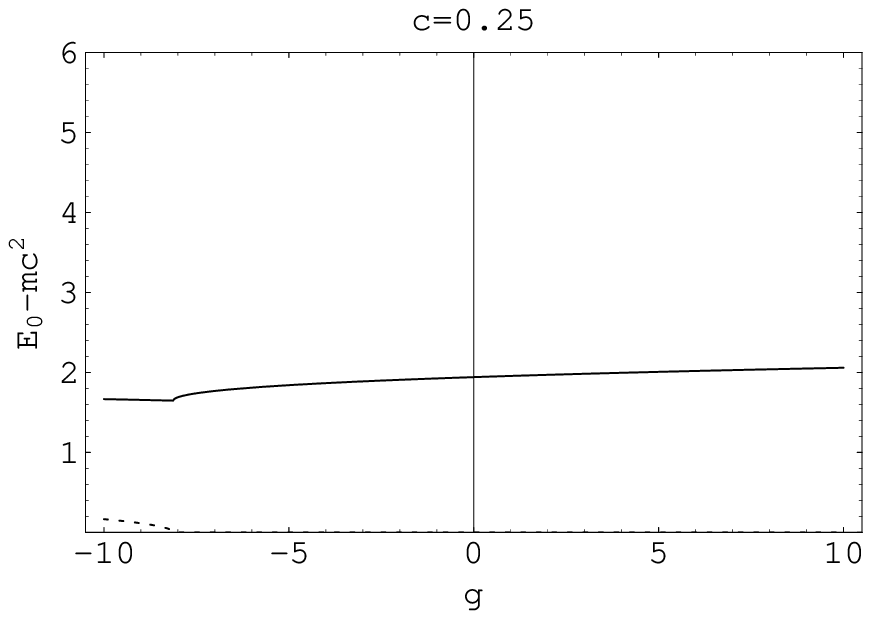,height=5.52cm,width=7.93cm}
\end{minipage}
\caption{\label{fig2} The behaviour of the ground state energy-level of the relativistic linear singular oscillator \eref{26} for values of speed of light $c= \infty$ (the non-relativistic case), $4$, $2$, $1$, $0.5$ and $0.25$ ($m=\omega=\hbar=1$). Real parts are shown by solid lines, imaginary parts by dashed lines. We see that, decreasing of $c$ changes appearance point of imaginary part ("collapse" point $g<-\frac{1}{8}\left(1+\frac{1}{4c^4}\right)$), too.}
\end{figure}
We note that, hermiticity condition of Hamiltonian $H$ imposes a restriction to the values of the quantity $g$. Indeed, from expression \eref{31a} it follows that $H$ is hermitian operator and its eigenvalues $E_n$ \eref{26} are real only in the case, when $\alpha$ and $\nu$ are real or complex-conjugate quantities. Therefore, $g$ must satisfy the condition $g> - \frac{\hbar^2}{8m}-\frac{\hbar^2 \omega_0^2}{32m}$. The case $g< - \frac{\hbar^2}{8m}-\frac{\hbar^2 \omega_0^2}{32m}$ ("collapse") will be considered separately.

Since the parameter $\mu=\frac{mc^2}{\hbar \omega} \rightarrow \infty$ when $c \rightarrow \infty$, by using \eref{a1} in Appendix A it is easy to show that in the limit case, when the velocity of the light approaches $\infty$, wavefunctions \eref{30} coincide with wavefunctions \eref{2} of the non-relativistic linear singular oscillator.

Energy spectrum \eref{26} also has a correct non-relativistic limit, i.e.

\begin{equation*}
E_n - mc ^2 \rightarrow E_n ^{nonrel} = \hbar \omega (2n+d+1) \; .
\end{equation*}

By the use of \eref{a2} it can be shown that, in the $g \rightarrow 0$ limit, the eigenfunctions \eref{30} and the eigenvalues \eref{26} transform into the "antisymmetrical" eigenfunctions and eigenvalues of the relativistic linear harmonic oscillator, considered in \cite{atakishiyev1}:
\begin{eqnarray}
\psi_n ^{relosc} (\rho) = c'_n \left[ \nu' (\nu'-1) \right] ^{-i \rho / 2} \Gamma( \rho + i \nu') P_{2n+1} ^{\nu'} \left( \rho ;  \frac \pi 2  \right), \nonumber \\
E_n ^ {relosc} =  \hbar \omega \left( 2n +1+ \nu' \right) \; , \quad n=0,1,2,3, \dots \; . \nonumber
\end{eqnarray}

Here $P_n ^{\nu'} \left( \rho ;  \varphi  \right)$ are the Meixner-Pollaczek polynomials and

\begin{equation*}
c'_n = 2^ {\nu'} \sqrt{ \frac{(2n+1)!}{2 \pi \lambda \Gamma(2n+1+ \nu')}}\;, \quad \nu' = \frac 1 2 + \sqrt{\frac 1 4 + \frac1 {\omega _0 ^2}}\; .
\end{equation*}

In \Fref{fig1} we show the behaviour of the ground state wave function $\psi_0(x)$ \eref{30} in dependence of the $x$ and $g$ for various values of the speed of light $c$. We see that except for the non-relativistic case, where $\psi_0(x)$ is real for $g \geq -\frac{\hbar^2}{8m}$, in the relativistic case ground state wavefunction is complex and has a lot of number of zeros. Similarly, in \Fref{fig2} we show the behaviour of the ground state energy-level \eref{26} in dependence of the $g$ for various values of the speed of light $c$. From these plots we see that decreasing of $c$ changes appearance point of imaginary part ("collapse" point $g<-\frac{\hbar^2}{8m}-\frac{\hbar^2 \omega_0^2}{32m}$).

\section{Conclusion}

The application of the finite-difference relativistic quantum mechanics to a large class of physical problems requires relativistic generalizations of the exactly solvable problems of non-relativistic quantum mechanics.

In this paper we construct a relativistic model of the linear singular oscillator and explicitly solve the corresponding finite-difference equation. We determine eigenvalues and eigenfunctions of the problem. They have correct non-relativistic limits. As in the non-relativistic case \cite{calogero}, the simplicity of the obtained energy spectrum suggests that a solution by group-theoretical methods should also be possible. Furthermore, we study some properties of the continuous dual Hahn polynomials.

We hope that, the model of the relativistic linear oscillator proposed in this paper will be applied in future in various fields of quantum physics likewise the non-relativistic singular oscillator.

\appendix
\section{}
\vskip 0.5cm

Here we present some formulas for the continuous dual Hahn polynomials \eref{28} used in the text.
\vskip 1.1cm
1. From the recurrence relation for the continuous dual Hahn polynomials \eref{29}, it can be shown that the following limit to the Laguerre polynomials holds:

\begin{equation}
\label{a1}
\lim _{\mu \rightarrow \infty} \frac 1 {n! \mu ^n} S_n \left(z \mu ; a,b, \frac12 \right) = L_n ^{a_0 - \frac 12} (z) \; , 
\end{equation}
where $a_0 = \lim \limits_{\mu \rightarrow \infty} a$ and $ \lim \limits_{\mu \rightarrow \infty} \left(b-\mu \right) = const$.
\vskip 1.1cm
2. It can be shown that, in some particular cases continuous dual Hahn polynomials coincide with the Meixner-Pollaczek polynomials, i.e.

\begin{equation}
\label{a2}
P_{2n+1} ^b \left( x; \frac \pi 2 \right) = (-1) ^n \frac{2^{2n+1}}{(2n+1)!} x S_n \left(x^2 ; 1, b, \frac 12\right)
\end{equation}
and

\begin{equation}
\label{a3}
P_{2n} ^b \left( x; \frac \pi 2 \right) = (-1) ^n \frac{2^{2n}}{(2n)!} S_n \left(x^2 ; 1, b, \frac 12\right) \; .
\end{equation}

To prove formulas \eref{a2} and \eref{a3} let's compare equations \cite{koekoek}

\begin{equation}
\label{a4}
\fl \quad \left[ e^{-i\varphi }\left( b+ix\right) e^{-i\partial _x}-e^{i\varphi }\left( b-ix\right) e^{i\partial _x}\right] y_1\left( x\right) =2i\left[ x\cos \varphi -\left( k+b\right) \sin \varphi \right] y_1\left( x\right)
\end{equation}
and

\begin{equation}
\label{a5}
\fl \quad \left[ \left( a+ix\right) \left( b+ix\right) e^{-i\partial _x}-\left( a-ix\right) \left( b-ix\right) e^{i\partial _x}\right] y_2\left( x\right) =2ix\left( 2n+a+b\right) y_2\left( x\right)
\end{equation}
where $y_1(x)=P_k^b \left(x, \varphi \right)$ are the Meixner-Pollaczek polynomials and $y_2(x)= S_n \left(x^2; a, b, \frac12 \right)$ are the continuous dual Hahn polynomials. It is easy to verify that, when $\varphi=\frac \pi 2$ and $k=2n+1$, Eq. \eref{a4} coincides with Eq. \eref{a5} for $a=1$. It means

\begin{equation}
\label{a6}
P_{2n+1}^b \left(x; \frac \pi 2 \right) = N_n x S_n \left( x^2; 1,b, \frac 12 \right) \; .
\end{equation}

Comparing coefficients, for example, for $x^{2n+1}$ in the left-hand and right-hand sides of \eref{a6} we find that,

\begin{equation*}
N_n = (-1) ^n \frac {2^{2n+1}}{(2n+1)!} \; .
\end{equation*}

One can prove relation \eref{a3} in same way.
\vskip 1.1cm
3. In the non-relativistic limit, when $\mu = \frac {mc^2}{\hbar \omega} \rightarrow \infty$ we have

\begin{eqnarray}
\label{a7}
\lim \limits_{\mu \rightarrow \infty }\alpha =d+\frac 12 \; , \nonumber\\
\lim \limits_{\mu \rightarrow \infty }\left( \nu -\mu \right) =\frac 12 \; , \nonumber \\
\lim \limits_{\mu \rightarrow \infty }\left( -\rho \right) ^{\left( \alpha \right) }=e^{\frac 12\left( d+\frac 12\right) \ln \mu }\left( -\xi \right) ^{d+\frac 12} \; , \\
\lim \limits_{\mu \rightarrow \infty }M\left( \rho \right) =\sqrt{2\pi }e^{\mu \ln \mu -\mu -\frac{\xi ^2}2} \; , \nonumber\\
\lim \limits_{\mu \rightarrow \infty }c_n=\frac 1{\pi \sqrt{n!\Gamma \left( n+d+1\right) }}e^{\mu -\left( \mu +n+\frac d2\right) \ln \mu } \; , \nonumber
\end{eqnarray}
where $\xi = \sqrt{\frac{m \omega}{\hbar}}x$. To obtain these formulas we used the representation

\begin{eqnarray}
\Gamma (z) \simeq \sqrt{\frac{2 \pi}{z}} e^{z \ln z - z} \; , \nonumber \\
|z| \rightarrow \infty \; , \nonumber
\end{eqnarray}
for the gamma function.
\vskip 1.1cm
4. In the non-relativistic limit we have following limit relations for operators \eref{32b} and \eref{32c}

\begin{eqnarray}
\lim \limits _{\mu \rightarrow \infty} \sqrt{\mu} a^- = c^- = - \frac{i}{\sqrt{2}} \left( \partial _\xi + \xi - \frac{d+ \frac 12}{\xi} \right) \; , \nonumber \\
\lim \limits _{\mu \rightarrow \infty} \sqrt{\mu} a^+ = c^+ = \frac{i}{\sqrt{2}} \left( -\partial _\xi + \xi - \frac{d+ \frac 12}{\xi} \right) \; . \nonumber
\end{eqnarray}

Hamiltonian \eref{31a} in this limit coincides with Hamiltonian of the non-relativistic singular oscillator \eref{1}:

\begin{equation*}
\lim \limits_{\mu \rightarrow \infty} H = H_N = \hbar \omega \left( c^+ c^- +d +1 \right) = \frac{\hbar \omega}2 \left( - {\partial _\xi}^2 + \xi ^2 + \frac{d^2 - \frac 14}{\xi^2} \right) \; .
\end{equation*}

\end{document}